\documentclass[%
 reprint,
 amsmath,amssymb,
prb,
]{revtex4-1}

\usepackage{graphicx}
\usepackage{dcolumn}
\usepackage{bm}

\begin{document}

\preprint{APS/123-QED}

\title{An Empirical Method to Measure Stochasticity and Multifractality in Nonlinear Time Series}%

\author{Chih-Hao Lin$^{1,2}$, Chia-Seng Chang$^{1,2}$, and Sai-Ping Li$^2$}
\affiliation{$^1$Department of Physics, National Taiwan University, Taipei 106, Taiwan}

\affiliation{$^2$Institute of Physics, Academia Sinica, Nankang, Taipei 115, Taiwan}

\date{\today}

\begin{abstract}
An empirical algorithm is used here to study the stochastic and multifractal nature of nonlinear time series.  A parameter can be defined to quantitatively measure the deviation of the time series from a Wiener process so that the stochasticity of different time series can be compared.  The local volatility of the time series under study can be constructed using this algorithm and the multifractal structure of the time series can be analyzed by using this local volatility.  As an example, we employ this method to analyze financial time series from different stock markets.  The result shows that while developed markets evolve very much like an Ito process, the emergent markets are far from efficient. Differences about the multifractal structures and leverage effects between developed and emergent markets are discussed.  The algorithm used here can be applied in a similar fashion to study time series of other complex systems.
\begin{description}
\item[PACS numbers] 
\pacs{}02.50.Fz, 05.45.Df, 05.45.Tp
\end{description}
\end{abstract}

\pacs{Valid PACS appear here}
\maketitle


\section{Introduction}	
Many complex systems in nature have time series datasets that exhibit non-stationary and complex behavior.  In many of these time series, their fluctuations show a wide range of time scales and/or broad distributions of the values.  These natural fluctuations are often found to follow a scaling relation over several orders of magnitude.  In general, such scaling laws could allow for a characterization of the data and the generating complex system by fractal (or multifractal) scaling exponents, which can reflect the characteristics of the systems in comparison with other systems and with different models.  Such fractal (or multifractal) scaling behavior has been observed in many time series from experimental physics, geophysics, physiology, etc.  Since these time series have many features in common, it would lead one to suggest a unified approach to study their behavior.  Many different methods have been developed to analyze these nonlinear time series.  In many of the studies on time series of these complex systems, researchers would assume they obey some stochastic processes, with stochastic differential equations of the type
\begin{eqnarray}
	dX=a(X,t)dt+b(X,t)dW_x, \label{eqn1}
\end{eqnarray}
where $X$ is the variable that we are interested in, $a(X,t)$ and $b(X,t)$ are variables that can depend on $X$ and $t$, the time and $W_x$ is a Wiener process. The time series generated is also assumed to have some long-range correlations.  We will take the above equation as our starting point.  One can indeed ask a straightforward but also fundamental question --- how close can the evolution of these nonlinear time series be described by stochastic processes that follow Eq.(\ref{eqn1})? From a practical point of view, one would just assume that the empirical time series follows exactly a stochastic process governed by Eq.(\ref{eqn1}). One would then decompose the empirical time series $dX$ into its components $a(X,t)$,  $b(X,t)$ and $dW_x$ as suggested by Eq.(\ref{eqn1}) by some optimization methods and to quantitatively measure its deviation from a stochastic process. One could further investigate the multifractal structures embedded in these time series.   From the computational point of view, this requires one to develop more efficient numerical algorithms in order to perform an effective optimization search for the resulting decomposed time series.  We will demonstrate this with a simple empirical algorithm in the present study and show how one can use the results obtained from the optimization search to investigate the interesting properties embedded in the time series.

\begin{figure}
\centering
\includegraphics[width=8cm]{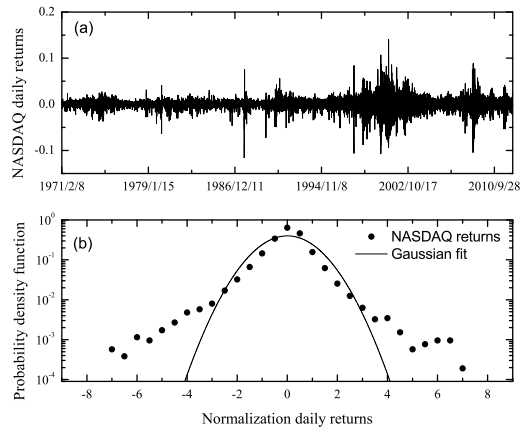} 
\caption{(a) a plot of the empirical data of the daily returns for the NASDAQ Composite index from 1971/2/8 $\sim$ 2012/6/13 (b) the probability density function of the normalized daily returns for the same period. }
\label{fig1}       
\end{figure}

Among the many nonlinear time series, of particular interest and also being intensively studied are financial time series.  Many established facts (or stylized facts as commonly called by practitioners in the field of economics and finance) in financial time series have already been identified and well documented\cite{ref19}.  These include e.g., the heavy tail distributions of asset returns and volatility clustering.  It is found that the heavy tail distributions of asset return exhibit power law behavior over several orders of magnitude.  Furthermore, the volatility exhibits long range correlations following an approximate power law while its fluctuations are randomly distributed and follow approximately a lognormal distribution\cite{ref18,ref18-1}. Such scaling behavior suggests that the time series might have a multifractal structure, which could reflect the characteristics of the systems in comparison with other systems and with various models.  These structures will certainly help researchers build better models to describe the underlying dynamics of financial markets and complex systems in general\cite{ref5-1,ref8,ref20}.  We will therefore choose the financial time series here as an example of our empirical study of nonlinear time series.  However, the empirical algorithm used here can also be applied to time series of other complex systems.  Indeed, it is known that time series of many complex systems reveal behaviors that are similar to the stylized facts observed in financial time series.  In financial markets, prices of stocks and commodities fluctuate over time, which then produce financial time series.  These time series are in fact of great interest both to practitioners and theoreticians for making inferences and predictions. Historically, it is noted that the well-known Efficient Market Hypothesis\cite{ref1,ref1-1} has a very close link to the stochasticity of the financial market.  In the case of financial time series, the stock prices $S_t$ at time $t$ are generally assumed to be an approximate Ito process given by the following form,
\begin{eqnarray}
	\frac{dS_t}{S_t} = \mu dt + \sigma dW_S, \label{eqn2}
\end{eqnarray}
where $\mu$ is the instantaneous risk free rate, $\sigma$ is the local volatility, and $W_s$ is a Wiener process, representing the inflow of randomness into the dynamics. The amplitude of this randomness is measured by the local volatility  $\sigma$ whose amplitude is usually related to risk in the market.  It should be noted that $\mu$ is a parameter that may depend on $S_t$,  $\sigma$ and $t$. When analyzing financial time series, researchers usually assume $\mu$ and $\sigma$ to take some constant values.  In reality, the amplitude of $\sigma$ is usually related to risk and its value can therefore vary as time evolves. Figure \ref{fig1}(a) is a plot of the empirical data of the daily returns for the NASDAQ Composite index from February 8, 1971 through June 13, 2012, while Figure \ref{fig1}(b) is the probability density function of the normalized daily returns for the same period. It is easy to see, that the distribution in Figure \ref{fig1}(b) has heavy tails on both ends and is far from a Gaussian distribution. If the fluctuations of the daily returns of the index obeyed the form as suggested by Eq.(\ref{eqn2}), $\sigma$ could not be a constant over the time period under study and should instead vary as a function of time.  It is noted that many time series of different complex systems also exhibit behavior similar to that of Figure \ref{fig1}(b), e.g., heart rate variability\cite{ref13}. Since these nonlinear time series have similar behavior, one would be tempted to develop a general framework that can be applied to time series of complex systems in general. 
Based on the assumption that financial time series should evolve very closely to stochastic processes, researchers also introduced the concept of stochastic volatility, assuming that the change of $\sigma$ also follows a stochastic process. Over the years, many stochastic volatility models\cite{ref2,ref2-1} have been proposed to evaluate derivative securities, such as options. For example, Hull and White\cite{ref2,ref2-1} proposed that the stochastic volatility obeys a stochastic differential equation of the type
\begin{eqnarray}
\begin{array}{clr}
\frac{dv_t}{v_t}&=\alpha dt + \beta dW_v  \\
v &= \sigma^2,
\end{array}
\label{eqn3} 
\end{eqnarray}
where $\alpha$ and $\beta$ are functions depending on $S_t$, $\sigma$ and $t$. $W_v$ is another Wiener process which can have correlation $\rho$ with $W_S$. In general, the actual time evolving process that the stochastic variance $v$ follows could be very complex.

For a time series of Eq.(\ref{eqn1}) to exhibit fractal behavior, the absolute moments of the change of its parameter $X$ should take the following scaling form\cite{ref5},
\begin{eqnarray}
\begin{array}{clr}
	M(q,T) &= E(|\delta _TX(t)|^q) \\&= E(|X(t+T)-X(t)|^q) \\
	&\sim K_qT^{f(q)}, 
\end{array}
\label{eqn4}
\end{eqnarray}
where $K_q$ is a prefactor which depends on $q$ and $f(q)$ is a function of $q$. $E(...)$ here stands for the expectation value of the subject in the bracket. From the multifractal theory, there should exist an $f(q)$ for each moment $q$, such that $M(q,T)$ scales with $T$, with the prefactor $K_q$ as the scale $T \rightarrow 0$.  When $f(q)$ is linear in $q$, only a single scaling exponent $H$ is needed and $f(q)=qH$. $X(t)$ is said to be monofractal. If $f(q)$ is a nonlinear function in $q$, $X(t)$ is then said to be multifractal.  Although the underlying causes of the observed scaling are often not known in detail, the fractal or multifractal characterization can be useful to model the corresponding time series, and to give predictions regarding extreme events or future behavior. 

In recent years, there exist a considerable number of papers devoted to the modeling of multifractality in nonlinear time series, pioneered by Mandelbrot and his collaborators\cite{ref4,ref4-1,ref4-2}. In their work, they developed a Multifractal Model of Asset Returns (MMAR), which is a continuous-time process that captures the heavy tails and long-memory volatility persistence that are exhibited by many financial time series.  It is constructed by combining a standard Brownian motion with a random time deformation process that is specified to be multifractal. Subsequently, other researchers have introduced other multifractal models, noticeably Bacry et.al.\cite{ref5,ref5-1}. In order to model asset return fluctuations, these authors introduced a multifractal random walk process to describe the behavior of financial time series which they call Multifractal Random Walk model (MRW).  It is a generalization of the Mandelbrot cascades to stationary, causal continuous cascades.  This model is able to describe many of the stylized facts of financial time series with satisfaction.  In its original form, MRW can simply be considered as a stochastic volatility model where the (log-) volatility memory has a peculiar logarithmic shape, a feature in agreement with empirical observation. This model receives much attention because of its simplicity.  There also exist analytic solutions for this model\cite{ref12} and one could therefore fit the set of parameters in the model to the empirical data in a more precise way. During the past decade, these multifractal models have been used to study time series of complex systems on various topics\cite{ref13,ref14,ref15}.  

Closely related to the question of stochasticity, one could also ask how well a stochastic volatility model, such as that of the type of Eq.(\ref{eqn3}) can model the nonlinear time series data under study.  In this case, one would need to decompose the empirical time series into the form as suggested by Eq.(\ref{eqn2}) and (\ref{eqn3}) and quantitatively measure its deviation from a stochastic process.  One would then be able to investigate the multifractality of the nonlinear time series as well as to study the detail of the local volatility $\sigma$ itself. We should also mention here that one should be aware of the fact that the apparent multifractal behavior in financial time series could also be a result of multiscaling in the time series\cite{ref3,ref8} and possibly other unknown factors.

In the following, we will describe how one can use a simple empirical optimization algorithm to quantitatively analyze nonlinear time series with the assumption that both the stock prices and volatility evolve according to some stochastic processes.  In this way, one can quantitatively measure the deviation from stochasticity of the system and compare it with other systems of similar nature.  Likewise, one can analyze the multifractal structure of the time series and compare with models.  One could also build models that would describe the behavior of the system under study.  In Section 2, we will introduce the empirical algorithm for the analysis of stochasticity of a system under study.  In Section 3, we present the result by using our method on simulated time series with different distributions of fluctuations.  Section 4 contains the results obtained from empirical data and Section 5 is the conclusion.

\section{Method}
We here assume that both the stock prices and the volatility evolve stochastically as the starting point.  We can also take Eq.(\ref{eqn1}) for the stock prices alone as the starting point if we do not want to put any assumption on $\sigma$. One can of course use other assumptions that one deems to be appropriate as the starting point for the time series under study.  In the study of financial time series, there are various definitions of the volatility used by researchers. In our case, since we want to use a consistent algorithm to evaluate the volatility $\sigma$ in Eq.(\ref{eqn2}) without assuming any particular form, we will therefore define it to be the local function such that the fluctuations of $lnS$ are a result of the product of this function and a distribution of fluctuations that should be as close to $dW_S$ as possible, i.e., a Wiener process in accordance with the assumption of Eq.(\ref{eqn2}). This will be useful as a general algorithm if one wants to study time series from different complex systems.  To make things simple, we here assume that $\mu$ is a constant throughout the time series under study.  We will then average over the whole time series and get the mean value of $lnS$. We then subtract this mean value from each $lnS_i$, where $i$ is the $i$-th data point (time step) of the time series, so that the mean, and effectively $\mu$ of the new time series is equal to zero. In the next step, we will look for the local volatility $\sigma$ at each time step of this new time series.  We will look for the local volatility $\sigma$ of the time series such that the asset return time series (after subtracting the mean of the whole time series) will be a product of this local volatility and a process that is as close to a Wiener process as possible. Deviation of the reconstructed $dW_S$ from a Wiener process would be regarded as a measure of the deviation of this time series from the stochastic process as defined by Eq.(\ref{eqn2}). To proceed, one can of course perform an optimization search that constrains $dW_S$ to be as close to a Wiener process as possible and numerically find the empirical local volatility for the time series. This is however very time consuming and inefficient.  Therefore we will do this in two steps. In the first step, we compute the local volatility $\sigma$ of the time series by using the moving window approach.  One then use this function as a starting point to perform an optimization search subject to the constraints that we mentioned above in order to obtain the local volatility function.  To carry out this procedure, a moving window with size $N$ is first chosen. We put the window on the first event of the asset returns sequence and calculate the volatility of the first $N$ events by adding the square of their values and dividing the sum by $N$. We then move the window to the second event and again calculate the volatility of the next $N$ events by adding the square of their values and divide the sum by $N$. We repeat the same procedure until we finish scanning through the whole asset return sequence.  As a matter of fact, one can also choose $N$ to be a varying number instead of a constant.  We should remark here that if one only performs a moving window average without further optimizing $dW_S$ against a Wiener process, the resulting distribution of the fluctuations of the volatility would be significantly different from a lognormal distribution. 

The next step is to carry out an optimization search to get the empirical local volatility function of the financial time series under study.  In order to perform an optimization search, one needs to introduce a cost function.  In general, if one carried out an optimization search on financial time series by using Eq.(\ref{eqn2}) alone, the resulting local volatility $\sigma$ would have a distribution of its fluctuations close to that of a lognormal distribution, a stylized fact that is already known. One can of course perform a more specific analysis, e.g., by optimizing with respect to a specific model or a class of models.  In this way, one could also investigate how well the models would describe the time series under study.  For our purpose, we will only assume in addition that the local volatility of the financial time series has fluctuations that follow a lognormal distribution function without further specification.  This is consistent with the stylized fact in financial time series that the fluctuations of the volatility obey an approximate lognormal distribution. For time series of other complex systems, one may need to introduce other constraints in the cost function.  In the optimization search here, we will therefore optimize $dW_S$ and $dln(\sigma)$ against Wiener processes in order to obtain the local volatility $\sigma$. The idea, as mentioned above is to minimize the sum of the deviation of $dln(\sigma)$ and $dW_S$ from Gaussian distributions in the cost function. One would then proceed to obtain the local volatility $\sigma$ by using some optimization algorithms to optimize this cost function.  In this paper, we use genetic algorithm (GA)\cite{ref9,ref9-1} to do the optimization search.  We here adopt the standard procedures of genetic algorithm to do the search, i.e., in each generation, we carry out the operations of mutation, crossover and selection. In our optimization search, we use a population size of 500 chromosomes, each with the length equivalent to the number of data points in the time series under study.  To begin with, we generate a configuration for each chromosome in the population.  We then carry out the genetic operators: crossover, mutation and selection. We now pick a number of pairs (in our study, we pick about 20\% of the population) of parent chromosomes for the crossover procedure.  We randomly pick a crossover point in each of the pair of parent chromosomes and make the crossover to produce a pair of baby chromosomes.  Next, we carry out the mutation operator.  For the mutation operator, we set a mutation rate (between 0 and 10\% in our tests) for the whole population and randomly choose the sites within the chromosomes to mutate.  We now have a population of chromosomes consisting of the chromosomes from the original population plus the new baby chromosomes from mutation and crossover.  We then calculate the value of the cost function for each of these chromosomes.  We next select a certain number of chromosomes (500 in this case) with better fitness for the next generation.  We will again carry out the same set of genetic operators (mutation, crossover and selection) in the new generation until the optimization search stops for some preset criteria.  The cost function $F$ (or fitness as commonly called in GA) used in our search is as follows

\begin{eqnarray}
\begin{array}{clr}
	F =&c\times area_{diff}(dW_s, Gaussian) \\
	&+ area_{diff}(dln(\sigma), Gaussian) , 
\end{array}
\label{eqn5}
\end{eqnarray}
where $c$ is a constant whose value one can choose and $area_{diff}(dW_s, Gaussian)$ is the difference of the area under the curve of the distribution of $dW_S$ and that of the Gaussian distribution.  In our optimization search, we choose $c$ to be between 1 and 2 and we have performed the search for 500 generations.

\begin{figure}
\centering
\includegraphics[width=8cm]{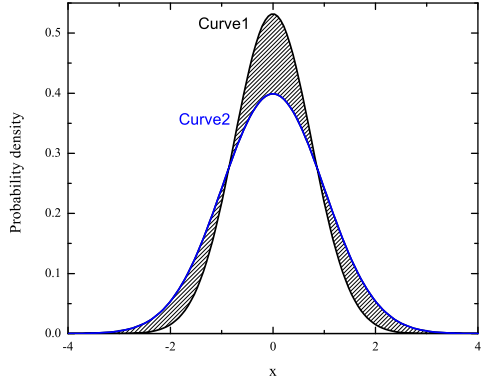} 
\caption{An illustrative diagram of $\Delta W_S$ in Eq.(\ref{eqn6}). The gray area is the non-overlapping region corresponding to the numerator in Eq.(\ref{eqn6}). }
\label{TwoGaussian}       
\end{figure}

From the optimization search, we now obtain the local volatility $\sigma$, its fluctuations $dln(\sigma)$ and the fluctuations of the asset return $dW_S$.  We can then plot the distribution of the fluctuations and compare with that of the distribution of a Wiener process. To put this on a more quantitative basis, one can define a quantity for each of the above distributions as the deviation of the distribution under study from a Wiener process.  Mathematically, a Wiener process has independent increments which will give a Gaussian distribution. We therefore define $\Delta W_S$ as the absolute value of the difference of the area under the curve of the probability density function (pdf) of $dW_S$ obtained from the optimization search and the curve of the pdf of a Wiener process, i.e., the non-overlapping region of the two distributions, divided by the sum of the area of the two.

\begin{eqnarray}
\begin{split}
	\Delta W_S &\equiv  \frac{A(|dW_S-Wiener|)}{A(dW_S)+A(Wiener)} \\
	&=\frac{A(|dW_S-Wiener|)}{2},
\end{split}
\label{eqn6}	
\end{eqnarray}
where $A(...)$ stands for the area of the subject in the bracket.  Figure \ref{TwoGaussian} is an illustrative diagram of Eq.(\ref{eqn6}).  Curves 1 and 2 are the pdf of $dW_S$ and a Wiener process respectively.  The gray area is the non-overlapping region of the pdf of the two distributions corresponding to the numerator in Eq.(\ref{eqn6}).  Since both $A(W_S)$ and $A(Wiener)$ are normalized to 1, the denominator is therefore equal to 2, as is shown in the last equality in the above equation.  If $dW_S$ is exactly equal to a Wiener process, there is perfect overlap and the difference is zero.  On the other hand, if the pdf of $dW_S$ and the Wiener process have no overlap at all, the difference as defined in the numerator of $\Delta W_S$  will be equal to the sum of the two distributions.  The parameter $\Delta W_S$ therefore has a value between 0 and 1.  The larger $\Delta W_S$ is, the more the time series deviates from a Wiener process.  In general, one believes that the time evolution of an efficient market should follow very closely to that of a Wiener process and therefore $\Delta W_S$ will be very small.  Large deviation from zero would indicate that the market is not totally efficient, or affected by other factors such as human psychology.  Using this parameter, one can compare different financial time series and also understand how efficient a financial market is.  This will then be a simple quantitative way for one to compare the efficiency of financial markets against each other and see how much they deviate from  a Wiener process.

In a similar fashion, one can define $\Delta dln(\sigma)$ as absolute value of the difference of the area under the curve of the pdf of $dln(\sigma)$ obtained from the optimization search and that of a Wiener process, divided by the sum of the area of the two.
\begin{eqnarray}
\begin{split}
	\Delta dln(\sigma) &\equiv \frac{A(|dln(\sigma)-Wiener|)}{A(dln(\sigma))+A(Wiener)}\\
	&=\frac{A(|dln(\sigma)-Wiener|)}{2}.
\end{split}
\label{eqn7}
\end{eqnarray}
With the reconstructed local volatility $\sigma$, its fluctuations $dln(\sigma)$ and the fluctuations of the asset return $dW_S$ in hand, one can further carry out analysis of other features and compare the result with the well-known stylized facts in financial markets.

\section{Simulated Time Series Test}

\begin{table*}
\caption{\label{table1} Result of the reconstructed $dW_S$ of simulated time series with different distributions of fluctuations by using our algorithm. }
\begin{ruledtabular}
\begin{tabular}{ccccc}
case & intrinsic $dW_S$ & KS test & reconstructed $dW_S$ &KS test\\ \hline
i    & 0.0\%     & T & 1.2\% & T \\
ii   & 19.8\%    & N & 17.8\% & N \\
iii  & 5.1\%	  & N & 5.3\% & N \\
iv 	 & 0.0\%     & T & 1.9\% & T \\
v    & 19.8\%	  & N & 12.8\% & N \\
vi   & 5.1\%	  & N & 4.3\% & N \\
vii   & 10.7\%	  & N & 6.1\% & N \\
\end{tabular}
\end{ruledtabular}
\end{table*}

In this section, we will test our method with some simulated time series of known distributions.  We will here use distributions including (i) Gaussian, (ii) rectangular, (iii) triangular, (iv) MRW, (v) MRW (with rectangular noise distribution), (vi) MRW (with triangular noise distribution), and (vii) MRW (with skew triangular noise distribution).  In the original version of MRW (case (iv)), the model consists of a lognormal distributed volatility function and a Brownian noise, whose fluctuations follow a Gaussian distribution.  In our tests, we also generate time series with other distributions as listed here.  It is a common practice to perform a null hypothesis test to compare whether two distributions are different from one another.  In order to check how well our algorithm performs, we here carry out a similar procedure by performing a Kolmogorov-Smirnov (KS) null test of the original simulated time series and our reconstructed time series in each of the cases listed above.  The KS test is applicable to unbinned distributions that are functions of a single independent variable and is known to be a useful approximation for null tests.  The result of the KS null test is included for comparison in Table \ref{table1} .  For the numerical test here, we generate 12,000 data points for each of the simulated time series, which is equivalent to a 40 $\sim$ 50 year period in financial markets.  In each of the cases (i) - (vii), we generated 100 time series and reconstruct the volatility and $dW_S$ by using our algorithm.  We then take their average.  Table \ref{table1} gives the results of the reconstructed time series of the simulated time series of different noise distributions by using our method.  In Table \ref{table1}, column one lists the noise distribution of the simulated time series.  Column two lists the original simulated time series with intrinsic deviation of the distribution of $dW_S$ from a Gaussian distribution as defined by Eq.(\ref{eqn6}).  Column three is the result of the KS test on these intrinsic distributions against a Gaussian distribution.  We set the significance level to be 0.01 in the test.  In the table, T means that one cannot distinguish the two distributions and N means one can.  At the significance level of 0.01, we can see that one cannot reject MRW + Gaussian but all the others fail the test.  The last two columns are the reconstructed $dW_S$ and the corresponding KS test.  One can see that the reconstructed time series using our method agree well with their corresponding simulated time series against the KS test.  The numerical test we carry out here thus show that our method will allow us to reconstruct a given time series to a good degree.  In the next section, we will apply our method to study the empirical time series from financial markets.

\section{Results}
The algorithm introduced above can now be used to analyze different financial time series.  In this paper, we have studied time series of Dow Jones Industrial Average (DJ, 1953/1/22$\sim$2012/6/28), S\&P500(1950/1/3$\sim$2012/6/13), Heng Seng Index (HSI, 1986/12/31$\sim$2012/6/20), London Stock Exchange (FTSE, 1984/4/2$\sim$2012/6/20), French Stock Market Index (CAC, 1990/3/1$\sim$2012/7/3), German Stock Index (DAX, 1990/11/26$\sim$2012/7/3), Shanghai Composite Index (SSE, 1990/12/19$\sim$2012/6/28), and Shenzhen Stock Exchange Composite Index (SZSE, 1991/4/3$\sim$2012/6/28). 

Let us now take the S\&P500 Index time series as an example here.  We here use the daily opening stock prices which means that we will take the time lag to be 1 day. Following Eq.(\ref{eqn2}), we first find $ln(\frac{S_{t+1}}{S_t})$ for the whole time series of the S\&P500 during the period January 3, 1950$\sim$June 13, 2012. We then calculate the mean $\mu$ of the series and subtract this mean from the sequence.  We next calculate the local volatility $\sigma$ of this new time sequence by using the algorithm introduced in the above. In our simulation, we use a window size between 20 and 30.  Our next step is to find the local volatility $\sigma$, its fluctuations $dln(\sigma)$ and the fluctuations of the asset return $dW_S$ of the S\&P500 by using the optimization algorithm discussed above.

\begin{figure}
\centering
\includegraphics[width=8cm]{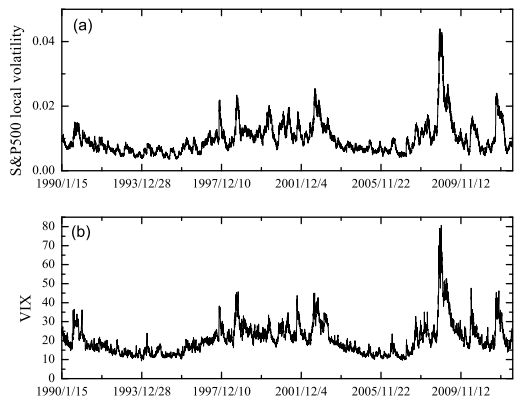} 
\caption{(a) The local volatility $\sigma$ obtained by our optimization algorithm during the period 1990 through 2012 (b) the VIX of the Chicago Board of Exchange (CBOE) during the same period. }
\label{fig2}       
\end{figure}

\begin{figure}
\centering
\includegraphics[width=10cm]{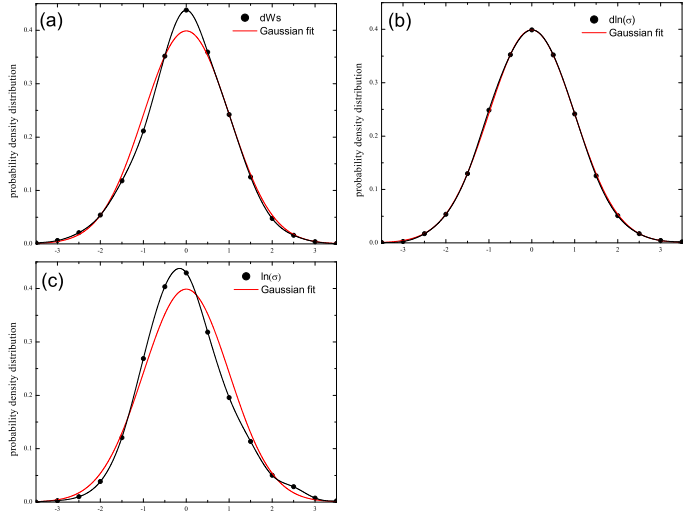} 
\caption{The probability distribution function(pdf) of (a)$dW_S$ (b)$dln(\sigma)$ (c) $ln(\sigma)$ of S\&P500. }
\label{fig3}       
\end{figure}

\begin{figure}
\centering
\includegraphics[width=8cm]{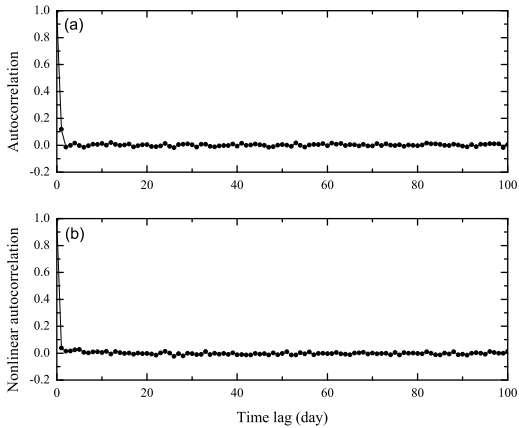} 
\caption{(a) The linear autocorrelation and, (b) the autocorrelation of the absolute value of $dW_S$ of S\&P500. }
\label{fig4}       
\end{figure}

\begin{figure}
\centering
\includegraphics[width=8cm]{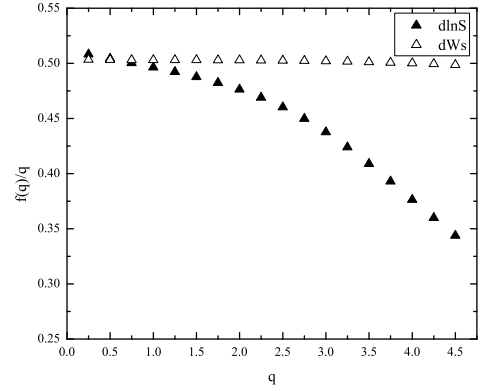} 
\caption{$f(q)$ vs. $q$ for both $dW_S$ and $dlnS$ of S\&P500.}
\label{fig5}       
\end{figure}

Figure \ref{fig2}(a) is the local volatility $\sigma$ obtained by our algorithm during the period 1990 through 2012.  For comparison, we show in Figure \ref{fig2}(b), the VIX of the Chicago Board of Exchange (CBOE). One can see that $\sigma$ thus obtained from the above optimization algorithm and that of the VIX in the Chicago Board of Exchange (CBOE) during the period of study have features very similar to each other.  The latter is sometimes also known as the Fear Index and is interpreted as the implied volatility of the S\&P500 Index by market practitioners. 
In a similar manner, we can now plot the distribution of the fluctuations $dln(\sigma)$ of the local volatility function and the fluctuations of the asset return $dW_S$ of the S\&P500 obtained from the optimization algorithm. The pdf of $dW_S$, $dln(\sigma)$ as well as $ln(\sigma)$ are shown in Figure \ref{fig3}.  The curves in red are the pdf of the standard normal distribution for comparison.  Note that since we want to compare the result with a standard normal distribution, we have normalized $dln(\sigma)$ by an overall factor $\sqrt{<dln(\sigma)^2>}$, which is the square root of the average of $dln(\sigma)^2$ of the whole time series.  In a similar fashion, for the result shown in Figure  \ref{fig3}(c), we have subtracted the average of $ln(\sigma)$ and then divided by the average of its standard deviation.  We see that both the pdf of $dW_S$ and $dln(\sigma)$ are very close to that of a Gaussian distribution. Its deviation, $\Delta W_S$ can be obtained as defined in Eq.(\ref{eqn6}). In the case of S\&P500 we study here, $\Delta W_S$ and $\Delta d ln(\sigma)$ are about 3\% and 0.6\% respectively, which are small deviations from a Gaussian distribution. Since the pdf of $dW_S$ and $dln(\sigma)$ are very close to that of a Wiener process, one can plot their corresponding linear and nonlinear autocorrelation functions.  In our study here, the linear autocorrelation of $dW_S$ is defined as $<dW_S(t+n) dW_S(t)>$, where $n$ is the time lag in days.  The nonlinear autocorrelation of $dW_S$ is defined as the autocorrelations that are different from the linear autocorrelation as defined above.  In this paper, we choose to study the autocorrelation of the absolute value of $dW_S$.  As an illustration, Figure \ref{fig4} shows the linear and nonlinear autocorrelation functions of $dW_S$. Figure \ref{fig4}(a) is the linear autocorrelation of  $dW_S$ while Figure \ref{fig4}(b) is the autocorrelation of its absolute value.  It is easy to see that they both resemble the behavior of that of a Gaussian noise time series.  The behavior of the linear and nonlinear autocorrelation functions of $dln(\sigma)$ is very similar that of Figure \ref{fig4}, suggesting that there are no significant long time correlations among the fluctuations in the volatility of the time series. Evaluation of the Hurst exponent $H$ of $dW_S$ indeed gives a value very close to 1/2, indicating that it follows very closely to that of a Wiener process. One can also use Eq.(\ref{eqn4}) to compute the moments of the temporal correlations of $dW_S$. The result in Figure \ref{fig5} indicates that $dW_S$ indeed has monofractal characteristic.  On the other hand, the dependence of the temporal correlations of $dlnS$ on $q$ suggests that it has multifractal structure.  The multifractal behavior could well be studied from the local volatility $\sigma$ itself.

\begin{figure}
\centering
\includegraphics[width=8cm]{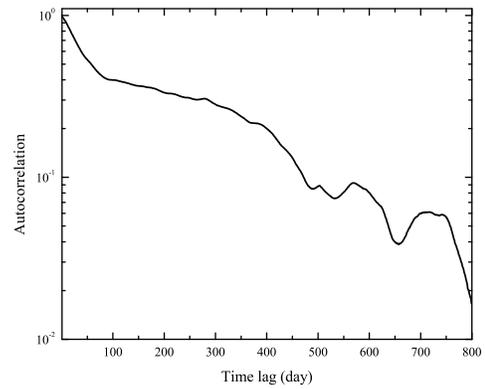} 
\caption{The autocorrelation function of the volatility $\sigma$ of S\&P500.}
\label{fig6}       
\end{figure}

The insignificance of the correlations among fluctuations in $dW_S$ and $dln(\sigma)$ can also be studied by other methods such as using the clustering index as introduced in \cite{ref6,ref7}, which demonstrates that they are indeed very close to that of the time series of a Gaussian noise. On the other hand, the autocorrelation function of the volatility $\sigma$ shows a slow decay and is shown in Figure \ref{fig6}, in agreement with the stylized facts observed in financial markets.

\begin{figure}
\centering
\includegraphics[width=8cm]{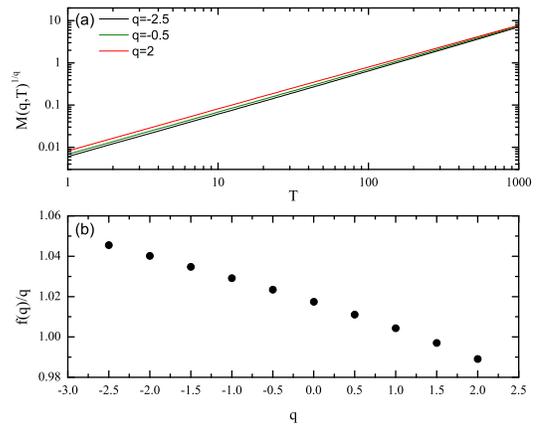} 
\caption{(a) shows $M(q,T)$ as a function of $T$ for different $q$ of $\sigma$ for S\&P500 and, (b) is its $f(q)$ as a function of $q$. The values fall on an approximate straight line within this range of $q$.}
\label{fig7}       
\end{figure}

Since $dW_S$ follows very closely a Wiener process, one can analyze the multifractal structure of the time series by directly studying the local volatility $\sigma$.  The resulting multifractal structure will therefore reflect the behavior of the long range correlations of the volatility.  We here use $\sigma$ to obtain $f(q)$ in Eq.(\ref{eqn4}).  The result is shown in Figure \ref{fig7}.  Figure \ref{fig7}(a) shows $M(q,T)$ as a function of $T$ for several different $q$ and Figure \ref{fig7}(b) is the plot of $f(q)$ as a function of $q$. By disentangling $dW_S$ from $\sigma$ when evaluating $M(q,T)$ and $f(q)$, one can understand better the multifractal behavior of $\sigma$ and the system itself.  One may then be able to construct a multifractal model that can describe the long range behavior of the system. 

The results of other financial time series that we have studied are summarized in Table \ref{table2}. In Table \ref{table2}, we include the $\Delta W_S$ and $\Delta dln(\sigma)$ of the financial time series that we study here.    For comparison, we also include the KS null test at significance level 0.01 for both the $\Delta W_S$ and $\Delta dln(\sigma)$ in Table \ref{table2}.  We can see that for the developed markets (S\&P500, DJ, CAC, DAX, FTSE and HSI), they all pass the KS test for both the $\Delta W_S$ and $\Delta dln(\sigma)$.  Their deviations from a Wiener process are indeed relatively small as can be seen from the numerical values of their $\Delta W_S$ and $\Delta dln(\sigma)$ as listed in Table \ref{table2}. For the developed markets, their Hurst exponents are all close to 1/2. Since the $f(q)$ for both $dW_S$ and $dlnS$ of these developed markets are quantitatively very similar to that of the S\&P500 (as shown in Figure \ref{fig5}), therefore we will not show here. It is interesting to note that the financial markets from mainland China show larger deviation from a Wiener process, as can be seen in Table \ref{table2}.  The KS null test also reveals this fact.  The larger deviations might be interpreted as a reflection of other factors that distinguish a financial time series from a strict stochastic process.  If this deviation is taken solely as from human factor, $\Delta W_S$ could then be interpreted as a measure of the degree of human effect on the market.  We should remark here that the $\Delta dln(\sigma)$ for the Shanghai and Shenzhen markets listed in Table \ref{table2} also pass the KS null test, suggesting that the fluctuations of their volatility can be viewed as following a stochastic process.  

\begin{figure}
\centering
\includegraphics[width=8cm]{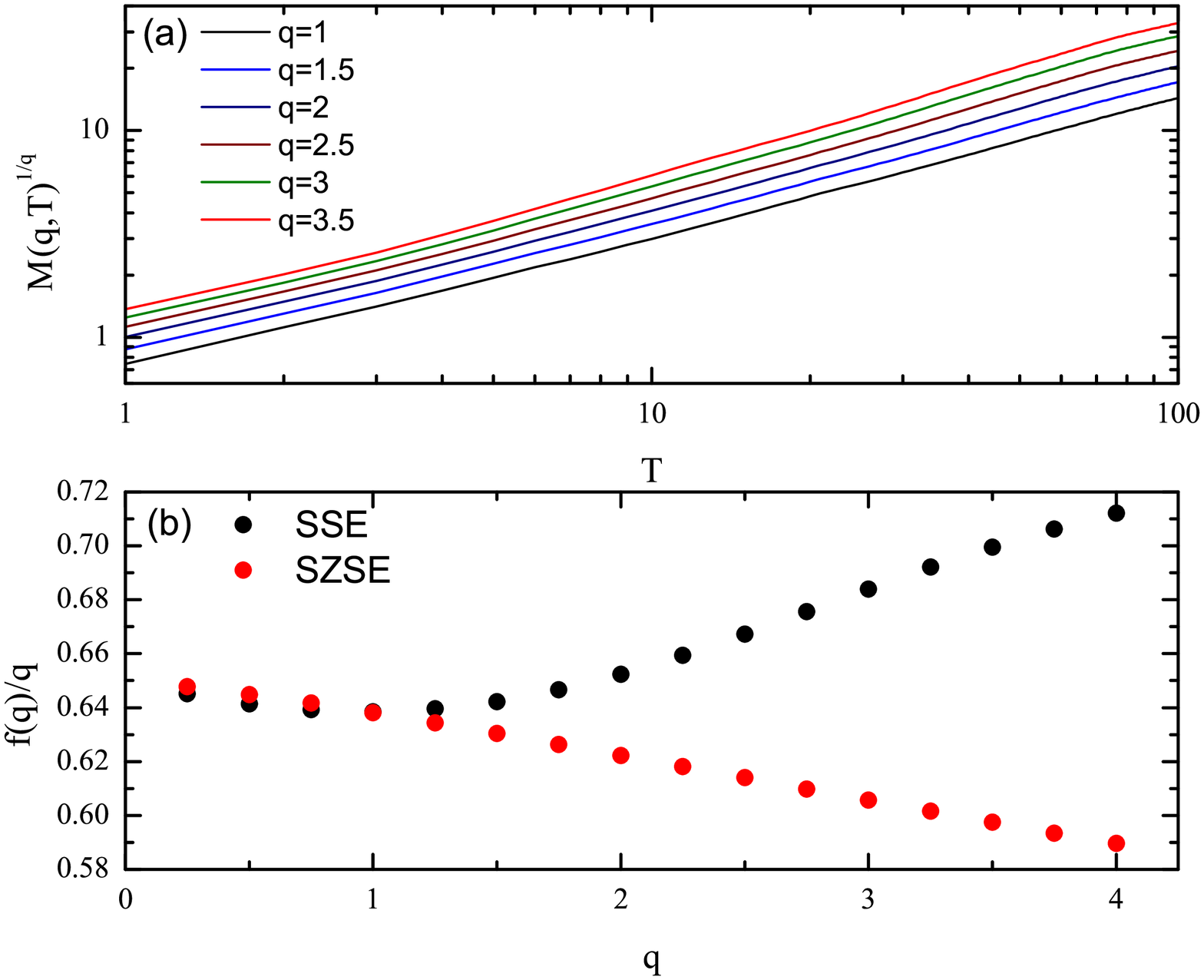} 
\caption{(a) shows $M(q,T)$ as a function of $T$ for different $q$ of $dW_S$ for Shanghai stock market index and, (b) is its $f(q)$ as a function of $q$.  Also shown is the $f(q)$ for the Shenzhen Stock Exchange Composite Index (red dots) for reference.}
\label{fig8}       
\end{figure}

Let us take a closer look at the Chinese stock markets.  The $dW_S$ of the Shanghai Composite Index is about 6\% from Table \ref{table2}.  A moment analysis of its $dW_S$ using Eq.(\ref{eqn4}) suggests that it has a multifractal behavior.  This is illustrated in Figure \ref{fig8}.  This result indicates that the Shanghai stock market cannot be described by a stochastic process as in Eq.(\ref{eqn1}).  Similarly, the Shenzhen Stock Exchange Composite Index also exhibits multifractal behavior for $dW_S$.  We therefore include the $f(q)$ of the reconstructed $dW_S$ of the Shenzhen Stock Exchange Composite Index in Figure \ref{fig8}(b) for reference.

\begin{table}
\caption{ \label{table2} $\Delta W_S$ and $\Delta dln(\sigma)$ of the financial time series studied here.} 
\centering
\begin{ruledtabular}
\begin{tabular}{ccccc}
& $\Delta W_S$ &KS test& $\Delta dln(\sigma)$ & KS test\\  \hline
S\&P500 &2.94\% &T  &0.55\% &T\\
DJ 		&2.37\% &T  &0.85\% &T\\
CAC		&2.16\% &T  &0.66\% &T\\
DAX		&3.30\% &T  &0.31\% &T\\
FTSE    &1.66\% &T  &0.51\% &T\\
HSI     &2.63\% &T  &1.03\% &T\\
SSE     &6.23\% &N  &1.33\% &T\\
SZSE    &5.00\% &N  &1.57\% &T\\
\end{tabular}
\end{ruledtabular}
\end{table}

\begin{figure}
\centering
\includegraphics[width=8cm]{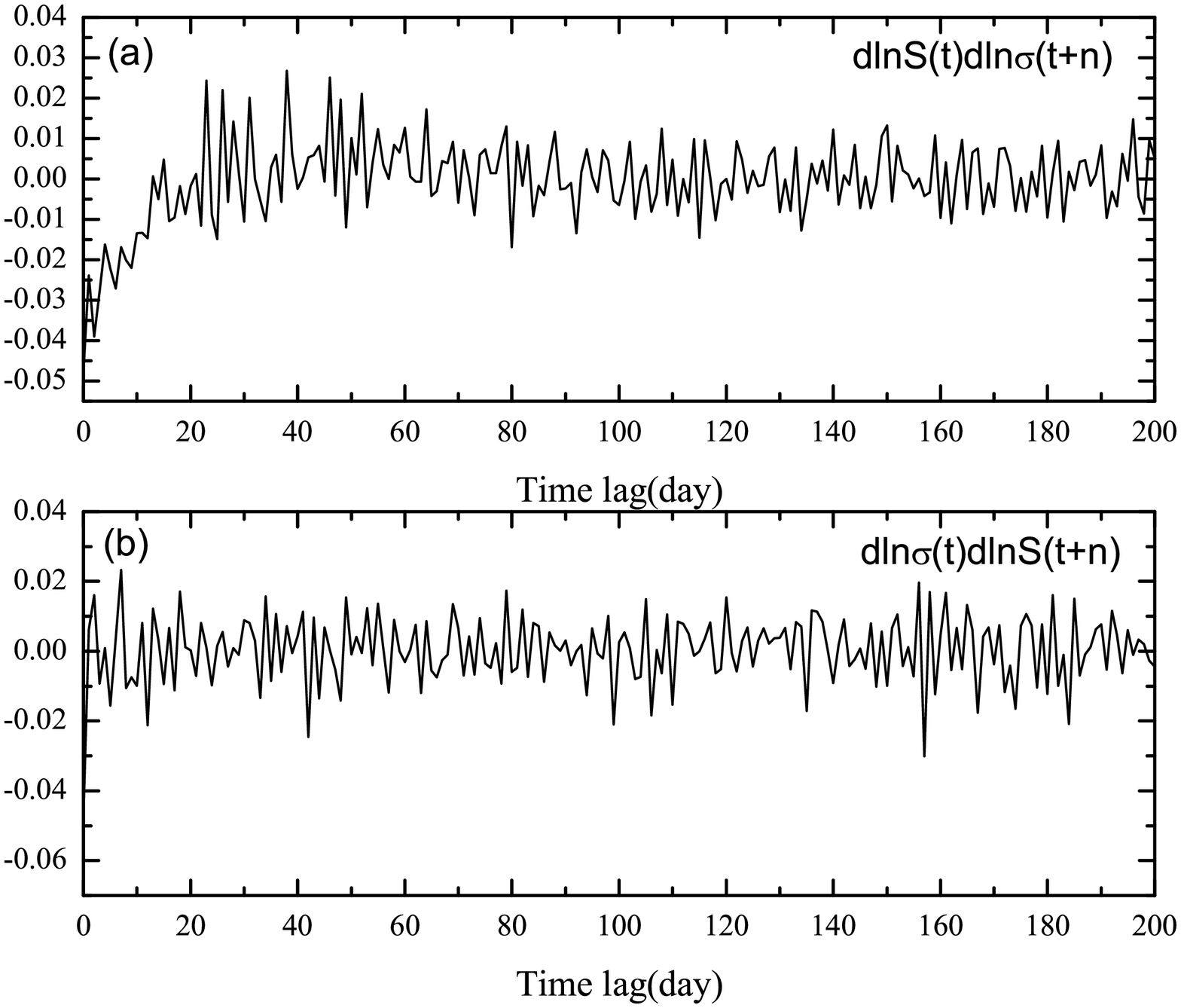} 
\caption{(a) the correlation of $<dlnS(t)dln(\sigma (t+n))>$ and, (b) the correlation of $<dln(\sigma (t))dlnS(t+n)>$ of the S\&P500.}
\label{fig9}       
\end{figure}

\begin{figure}
\centering
\includegraphics[width=8cm]{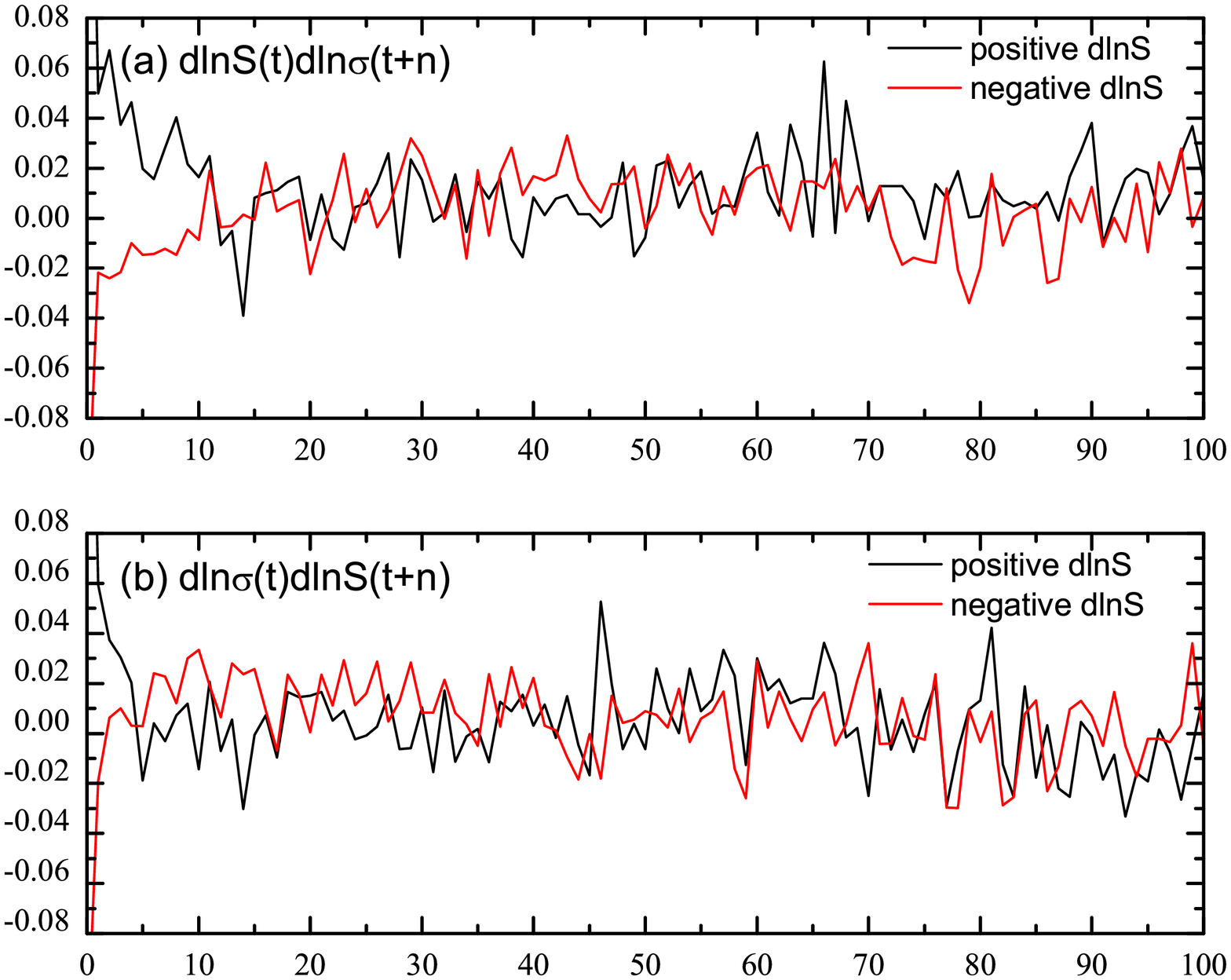} 
\caption{(a) the correlation of $<dlnS(t)dln(\sigma (t+n))>$ and, (b) the correlation of $<dln(\sigma (t))dlnS(t+n)>$ of the Shanghai index.}
\label{fig10}       
\end{figure}

It is known that negative price returns induce increased future volatilities, known as the leverage effect\cite{ref10} .  This means that the variation of the log-return in the past is found to be negatively correlated with the variation of the volatility in the future.  On the other hand, it is also known that the variation of the volatility in the past is not correlated with the variation of the log-return in the future\cite{ref11,ref11-1}. Putting this into mathematical language, it implies that there should be a correlation between $dln(S)$ and $dln(\sigma)$, or between $dW_S$ and $dln(\sigma)$.  More specifically, there should be a correlation between the large negative fluctuations of $dln(S)$ (or $dW_S$) at time $t$, and large positive fluctuations of $dln(\sigma)$, at time $t + n$, where $n$ is a positive integer, for the time series under study. Figure \ref{fig9}(a) shows the correlation of $<dlnS(t)dln(\sigma (t+n))>$ of the S\&P500.  One can see that there is a correlation between the large negative fluctuations of $dln(S)$ (or $dW_S$) at time $t$, and large positive fluctuations of $dln(\sigma)$, at time $t + n$.  No such correlation is observed for large positive fluctuations of $dln(S)$ (or $dW_S$) at time $t$, and large positive fluctuations of $dln(\sigma)$, at time $t + n$. The result shown in Figure \ref{fig9}(a) indicates that this correlation would last for about 10$\sim$15 trading days.  In a similar fashion, one can study the correlation of $<dln(\sigma (t))dlnS(t+n)>$ of the S\&P500 and the result is shown in Figure \ref{fig9}(b).  The correlations $<dW_S (t))dln(\sigma (t+n)))>$ and $<dln(\sigma (t))dW_S (t+n)>$ are very similar to that of the correlations in 
Figure \ref{fig9}(a) and \ref{fig9}(b) respectively and will not be shown here.

The results of the correlation of $<dln(S(t))dln(\sigma (t+n))>$ of other developed markets (DJ, CAC, DAX, FTSE and HSI) have behaviors very similar to that of S\&P 500 and therefore we do not show here.  For the financial indices in China, both large negative and positive fluctuations of $dln(S)$ at time t will induce large positive fluctuations of $dln(\sigma)$ at time $t + n$ and the result is shown in Figure \ref{fig10}.  The Shenzhen Stock Exchange Composite Index exhibits similar behavior for both large negative and positive fluctuations of $dln(S)$ to that of the Shanghai Composite Index which we do not show here.  This is another indication that the emergent markets evolve very differently from developed markets. In a similar fashion, one can study the high frequency financial time series data using this empirical method.  It should be noted that a daily detrending is needed in order to get a consistent result.  Analysis of stochasticity and multifractality of the high frequency financial time series would then follow the same procedures which we will not do here.

\section{Conclusion}
In this paper, we have employed a simple empirical method to study nonlinear time series of complex systems.  This is based on an optimization search of the local volatility $\sigma$ with the assumption that the time series follows closely an Ito process.  We here choose financial time series to illustrate this method with the assumption that the fluctuations of the log asset returns and the fluctuations of the log volatility follow very closely to that of a Wiener process. We propose parameters $\Delta W_S$ and $\Delta dln(\sigma)$ to quantitatively measure the deviation of a financial time series from a Wiener process. If these parameters are very close to zero, then the time series under study should follow very closely a stochastic process.  The deviation could be interpreted as an indication of the collective effect of factors such as human psychology that renders the time series to differ from a stochastic process.  By doing so, one is able to compare different financial markets and study how these factors could affect different markets in a more quantitative way.  We have studied several market indices in this paper. Using this algorithm, we found that the developed markets (DJ, S\&P500, CAC, DAX, FTSE and HSI) have relatively small $\Delta W_S$ values while the emergent markets (Shanghai Composite (SSE) and Shenzhen Stock Exchange (SZSE) Indices) exhibit larger $\Delta W_S$ values.  This reveals the fact that the developed markets in the US and Europe are closer to an efficient market than the emerging new markets such as those in China.  For the time series of the Chinese stock markets, the values of $\Delta W_S$ show a larger deviation from that of a Wiener process, meaning that the time series do not follow a stochastic process.  This is also supported by the KS null test.  This deviation indeed induces multifractal structures in $dW_S$.  It further gives leverage effects that are significantly different from those observed in the developed markets.  Whether these are the result of the non-stochasticity behavior of the time series is unclear and is worth to investigate in the future.   

While the nonlinear autocorrelation functions of the asset returns time series of the developed markets display power law like decay, the corresponding nonlinear autocorrelation functions of the $dW_S$ exhibit behaviors very similar to Gaussian noise time series, meaning that the $dW_S$ are indeed very close to a Wiener process. The more interesting results are the autocorrelation functions of the local volatility $\sigma$ of the time series and its fluctuations. The autocorrelation of the local volatility shows slow decay behavior, but both the linear and nonlinear autocorrelations of its fluctuations $dln(\sigma)$ show behavior very similar to a stochastic time series. If $dW_S$ for a time series is very close to a Wiener process, one could investigate the multifractal structure of the time series directly from the reconstructed local volatility obtained, for example, by the algorithm introduced here.  The local volatility $\sigma$ obtained from the time series under study might in fact be interpreted as a response function to the external and internal shocks that act on the system with its fluctuations follow closely to a lognormal distribution.

As mentioned above, the method proposed here to find the local volatility of a time series can be improved by developing better optimization algorithms than the one being used here.  Introducing appropriate constraints related to the time series under study into the cost function can further improve its performance.  The interested reader can design his own optimization algorithm to further improve the search performance on the local volatility $\sigma$ that can minimize the cost function under study.

Quantitative methods that are developed to study financial time series can indeed be used to study other complex systems\cite{ref16} as well.  The algorithm introduced here can be applied to analyze time series of other complex systems and thus to quantitatively measure their deviations from a Wiener process.  The stochasticity and structure of multifractality of these complex systems can then be studied in a similar fashion.  This empirical algorithm has recently been applied to the study of heart rate variability in cardiology and will be reported in a separate publication \cite{ref17}.  



\end{document}